\documentclass[%
 aip,
 jmp,%
 amsmath,amssymb,
 reprint,%
]{revtex4-1}

\usepackage{graphicx}% Include figure files
\usepackage{dcolumn}% Align table columns on decimal point
\usepackage{bm}% bold math
\usepackage{hyperref}
\usepackage{xcolor}
\hypersetup{
    colorlinks=true,
    linkcolor=blue,
    citecolor=blue,
    urlcolor=blue,
}
\usepackage{color,soul}

\begin{document}

%\preprint{AIP/123-QED}

\title{Tuning the magnetodynamic properties of all-perpendicular spin valves using He$^+$ irradiation}% Force line breaks with \\
%\thanks{Footnote to title of article.}

\author{S. Jiang}
\affiliation
{Department of Applied Physics, School of Engineering Sciences, KTH Royal Institute of Technology, Electrum 229, SE-16440 Kista, Sweden}
\affiliation{NanOsc AB, Kista 164 40, Sweden}

\author{S. Chung}
\affiliation
{Department of Applied Physics, School of Engineering Sciences, KTH Royal Institute of Technology, Electrum 229, SE-16440 Kista, Sweden}
\affiliation
{Department of Physics and Astronomy, Uppsala University, 751 20 Uppsala, Sweden}

\author{L. Herrera Diez}
\affiliation
{Institut d'Electronique Fondamentale, CNRS, Universit\'e Paris-Sud, Universit\'e Paris-Saclay, 91405 Orsay, France}

\author{T. Q. Le}
\affiliation{Department of Applied Physics, School of Engineering Sciences, KTH Royal Institute of Technology, Electrum 229, SE-16440 Kista, Sweden}
\affiliation{Department of Physics, University of Gothenburg, 412 96, Gothenburg, Sweden}

\author{F. Magnusson}%
\affiliation
{NanOsc AB, Kista 164 40, Sweden}

\author{D. Ravelosona}
\affiliation
{Institut d'Electronique Fondamentale, CNRS, Universit\'e Paris-Sud, Universit\'e Paris-Saclay, 91405 Orsay, France}
\affiliation
{Spin-Ion Technologies, 28 rue du general Leclerc, 78000 Versailles Cedex, France}

\author{J. \AA{}kerman}
 \altaffiliation[Author to ]{akerman1@kth.se.}
\affiliation
{Department of Applied Physics, School of Engineering Sciences, KTH Royal Institute of Technology, Electrum 229, SE-16440 Kista, Sweden}
\affiliation
{NanOsc AB, Kista 164 40, Sweden}
\affiliation
{Department of Physics, University of Gothenburg, 412 96, Gothenburg, Sweden}

\date{\today}% It is always \today, today,

\begin{abstract}
Using He$^+$ ion irradiation, we demonstrate how the magnetodynamic properties of both ferromagnetic layers in all-perpendicular [Co/Pd]/Cu/[Co/Ni] spin valves can be tuned by varying the He$^+$ ion fluence. As the perpendicular magnetic anisotropy of both layers is gradually reduced by the %He$^+$ ion 
irradiation, different magnetic configurations can be achieved from all-perpendicular ($\uparrow\uparrow$), through orthogonal ($\rightarrow\uparrow$), to all in-plane ($\rightrightarrows$). In addition, both the magnetic damping ($\alpha$) and the inhomogeneous broadening ($\Delta H_{0}$) of the Co/Ni layer improve substantially with increasing fluence. GMR of the spin valve is negatively affected and decreases linearly from an original value of 1.14\% to 0.4\% at the maximum fluence of 50$\times10^{14}$~He$^{+}$/cm$^{2}$.
%
%Valid PACS numbers may be entered using the \verb+\pacs{#1}+ command.
\end{abstract}

\pacs{Valid PACS appear here}% PACS, the Physics and Astronomy
\maketitle

%\section{\label{sec:level1}Introduction}

Pseudo-spin-valves (PSVs) are multilayer stacks consisting of two ferromagnetic (FM) layers separated by a nonmagnetic (NM) metallic spacer. They have received a great deal of attention %recently 
due to their applications in spintronics, which include magnetic read heads, sensors, and magnetoresistive random access memory (MRAM)\cite{Akerman2005,Engel2005,Bhatti2017MatTod}. %These are based on the giant magnetoresistive (GMR) effect,\cite{BinaschG, BaibichM1988} which relies on the relative magnetization configuration of two FM layers. 
More recently, spin-transfer torque (STT) MRAM\cite{Locatelli2014, Chappert2007} and spin-torque nano-oscillators (STNOs)\cite{SIKiselev, Rippard2004} have been investigated; these utilize the STT effect, which %. The STT effect 
describes how a current of spin-polarized charges can exert a torque on the magnetization of a ferromagnetic layer,  driving that layer into precession or switching.\cite{Berger1996, Slonczewski1996, Slonczewski1999a}
%These spin-valve-based devices can be classified into two categories based on their working principles. One is based on the giant magnetoresistive (GMR) effect\cite{BinaschG, BaibichM1988}, where the magnetoresistive (MR) relies on the magnetization configuration of two FM layers. This effect has been used to the commercialized magnetic read head for hard disc drivers\cite{Chappert2007} and magnetoresistive random-access memory (MRAM)\cite{Kudo2015}. The other is so-called spin-transfer torque (STT)\cite{Berger1996, Slonczewski1996}, where the charge current first polarized by one of the FM layer (fixed layer) and then exerts a torque on another FM layer (free layer). This STT hence can be used to manipulate the magnetic moments, either switch or processing. STT-MRAM\cite{Locatelli2014, Chappert2007} and spin-torque nano-oscillators (STNOs)\cite{SIKiselev, Rippard2004} are two typical applications. Based on the specific applications, different magnetic configurations are required and widely investigated. 

In early studies, STNOs were generally fabricated from spin valves with easy-plane magnetic anisotropy %(all-IMA) 
materials, such as NiFe and Co;  interesting dynamics of propagating spin wave and localized bullet solitons were reported.\cite{Bonetti2010b, Dumas2013a} A novel nano-scopic and magnetodynamical object---the magnetic droplet soliton---was then unveiled in orthogonal STNOs with perpendicular magnetic anisotropy (PMA) free layer and easy-plane fixed layer.\cite{Hoefer2010, Mohseni2011, Mohseni2013, Mohseni2013b, Chung2014, Iacocca2014, Macia2014, Lendinez2015, Chung2016, Xiao2016, Xiao2016a, Lendinez2017prapplied} Following this trend, stabilized droplets were found in all-PMA spin valve STNOs,\cite{Chung2017} thus enabling further insights into droplet dynamics, such as its size and nucleation boundary. Obviously, tailored PMA not only plays a crucial role in stabilizing magnetic droplets, it is increasingly important for future generations of STT-MRAM.\cite{Thomas2014, Bhatti2017MatTod} The engineering of PMA thus attracts much interest. For example, irradiation with ions such as  Ar$^{+}$, He$^{+}$, or Ga$^{+}$ has proven to be capable of  modifying the magnetic properties of multilayers, the PMA of which is  sensitive to surface or interface structures.\cite{Chappert1998, Rettner2002, Beaujour2009, Stanescu2008, Beaujour2011}

%Taking the STNOs as an example, in the early stages, both FM layers with in-plane magnetized (referred to all-IMA), such as mostly used NiFe and Co, have been explored extensively, since This all-IMA based STNOs are able to generate propagating spin-waves and localized bullets\cite{Dumas2013a, Bonetti2010b}. Then, orthogonal PSVs with PMA free layer (such as Co/Pt and Co/Ni Co/Pd multilayers) and easy-plane fixed layer are studied. A novel nanomagnetic structure - dynamical droplets - are unveiled.\cite{Mohseni2013, Macia2014, Chung2016} Their fundamental physics interests and inherent high emission power for microwave generators attract significant attentions. Very recently, we found this droplet become super stable in all-PMA STNOs\cite{Chung2017}, where both free and fixed layer are with strong PMA. These different architectures of STNOs require careful design of FM layers before the micro and nanofabrications. On the other hand, their magnetic properties various or are even damaged after experiencing the processing. It would, therefore, be highly necessary to control the magnetic properties, such as PMA, after film deposition or even in the end of the nanofabrication. Quite recently, ion irradiation, such as Ar$^{+}$, He$^{+}$ or Ga$^{+}$, has shown the capability to modify the magnetic properties of multilayers, whose PMA are sensitive to the surface and/or interface structures\cite{Chappert1998, Rettner2002, Beaujour2009, Stanescu2008, Beaujour2011}.

In this paper, we tailor the magnetodynamic properties in an all-PMA [Co/Ni]/Cu/[Co/Pd] PSV using  He$^{+}$ irradiation. We achieve a range of magnetic configurations, from all-perpendicular ($\uparrow\uparrow$), through orthogonal ($\rightarrow\uparrow$), to all in-plane ($\rightrightarrows$) by precisely controlling the He$^{+}$ fluences. Hysteresis loop (HL), magnetoresistance (MR), and broadband ferromagnetic resonance (FMR) measurements are performed. We find that the coercivity $H_{c}$, GMR, and effective magnetization $M_{\mathit{eff}}$ of both the Co/Ni and the Co/Pd layers can be controlled consistently. At the same time, the damping and inhomogeneous broadening of the Co/Ni layer show rather dramatic improvements, while those of Co/Pd seem unaffected. These observations are likely to prove essential to spintronic applications.

%\section{\label{sec:level1}Experimental Details}

Thin-film deposition was performed using a commercial AJA ATC Orion 8 sputtering system. A full stack of seed layer Ta~(5)/Cu~(15)/Ta~(5)/Pd (3), all-perpendicular spin valve [Co~(0.5)/Pd~(1.0)]${\times5}$/ Co~(0.5)/Cu~(6)/[Co~(0.3)/Ni~(0.9)]${\times4}$/Co~(0.3), and capping layer Cu~(3)/Pd~(3) was deposited on thermally oxidized Si substrate (numbers in parentheses are thicknesses in nanometers). The sample was then cleaved into nine pieces of about 7$\times $7~mm for He$^{+}$ irradiation. Using He$^{+}$ ions at energies around 15 keV, recoils are limited to 1-2 atomic distances and all the ions stop deep into the substrate. As a result, He$^{+}$ irradiation is a soft process that allows a precise control of interface intermixing maintaining both the crystalline texture and grain size of the pristine films. The fluence ($F$) of He$^{+}$ ions was varied from 2 to 50$\times10^{14}$~He$^{+}$/cm$^{2}$. Hysteresis loop measurements, where the applied field is either along (in-plane, IP) or normal (out-of-plane, OOP) to the film plane, were conducted using an alternating gradient magnetometer (AGM). To measure GMR, a commercial Picoprobe with G- and S-pads was applied to the extended films. The dc current was then injected into the film with a Keithley 6221 current source, and dc voltage was picked up by a Keithley 2182 nanovoltmeter. The broadband ferromagnetic resonance measurement was carried out using a NanOsc Instruments PhaseFMR-40 with a coplanar waveguide (CPW). The FMR frequency $f$ was between 3 and 40 GHz. An OOP magnetic field $H$ was swept from 0 to 1.3~T at a given frequency. All of the measurements were performed at room temperature.

%\section{\label{sec:level1}Results and Discussion}
\begin{figure}
\centering
\includegraphics[width=3.4in]{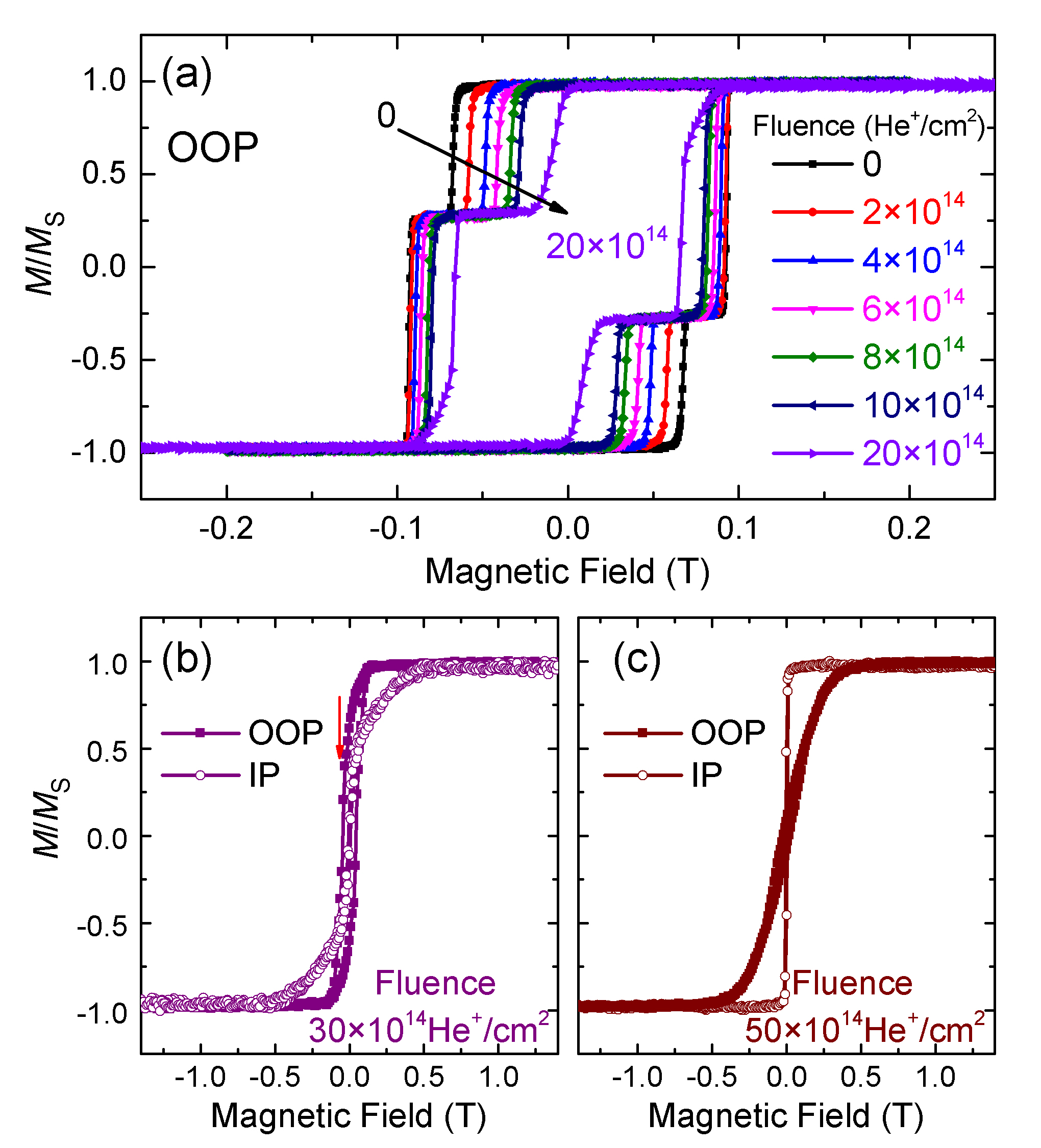}
\caption{\label{fig1} (a) Out-of-plane (OOP) hysteresis loops vs.~He$^{+}$ irradiation fluences ranging from 0 to 20$\times $10$^{14}$~He$^{+}$/cm$^{2}$. (b) and (c) in-plane (IP) and OOP hysteresis loops for fluences of 30 and 50$\times $10$^{14}$~He$^{+}$/cm$^{2}$.
}
\end{figure}

Figure~\ref{fig1}(a) shows the OOP hysteresis loops for different fluences. The nonirradiated film ($F$~=~0) shows clear two-step switching; the higher ($H_{c}$~=~93~mT) and lower ($H_{c}$~=~67~mT) coercivities correspond to Co/Pd and Co/Ni layers, respectively, indicating strong PMA. As  He$^{+}$ fluences increase, $H_{c}$  undergoes  continuous reduction and the squareness values remain approximate  up to $F$~=~10$\times$10$^{14}$~He$^{+}$/cm$^{2}$ for both Co/Ni and Co/Pd, suggesting that their PMA values were modified by He$^{+}$ irradiation. Further quantitative study on  PMA will be carried out by FMR later. As for higher fluences ($F$~=~30$\times$10$^{14}$~He$^{+}$/cm$^{2}$), Figure~\ref{fig1}(b) shows that both OOP and IP HLs show a continuous change, followed by a sharp drop in magnetization, indicating that one FM layer remains OOP magnetized while the other layer favors easy-plane---that is,  an orthogonal spin-valve. Figure~\ref{fig1}(c) shows that the saturation field for IP is much lower than that of OOP, which means that the remanent magnetic states of both Co/Ni and Co/Pd turn to in-plane magnetized %(all-IMA) 
for the highest fluence ($F$~=~50$\times$10$^{14}$~He$^{+}$/cm$^{2}$).

\begin{figure}
\centering
\includegraphics[width=3.4in]{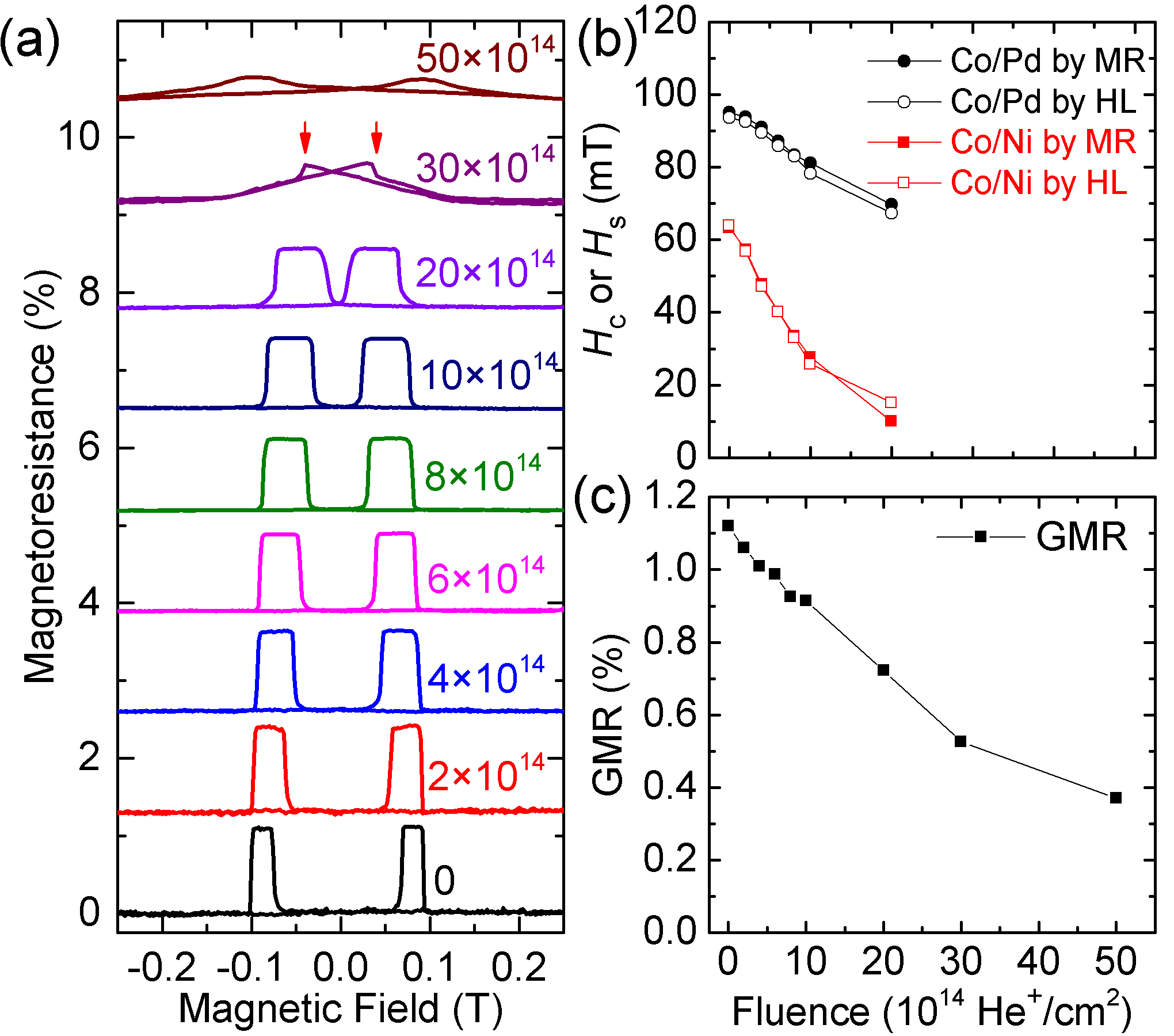}
\caption{\label{fig2} (a) Magnetoresistance measurements (MR) in an OOP field. (b) Coercivity $H_{c}$ and switching field $H_s$ extracted from MR and hysteresis loop (HL) measurements. (c) Extracted GMR value as a function of He$^{+}$ fluence. 
}
\end{figure}

In the following, we carried out MR measurements on all samples. The MR data are plotted in Fig.~\ref{fig2}(a). It is clear that the switching fields $H_{s}$ show similar decreasing trends as the HL data in Fig.~\ref{fig1}. We compared the $H_{c}$ and $H_{s}$ from both the HL and MR measurements, shown in Fig.~\ref{fig2}(b). As expected, $H_{c}$ and $H_{s}$ are in good agreement with each other. It should be noted that, for $F$~=~30 and~50$\times$10$^{14}$~He$^{+}$/cm$^{2}$, the overlapping signals of both Co/Ni and Co/Pd prevented the separation of $H_{c}$ and $H_{s}$; they were thus not extracted. The MR curve in Fig.~\ref{fig2}(a) for 30$\times10^{14}$~He$^{+}$/cm$^{2}$ shows  a clear MR jump at $H$~=~40~mT, which is consistent with the $H_{c}$ of Co/Pd at the same fluence as the red arrow in Fig.~\ref{fig1}(b). Both HL and MR curves are saturated at $H$~=~140~mT, which results from the saturation of the IP magnetized Co/Ni layer. In Fig.~\ref{fig2}(c), we calculated the GMR values defined by GMR~=~[$R$(AP)-$R$(P)]/$R$(P).\cite{BinaschG, BaibichM1988} These show a linear decreasing trend from 1.14\% to 0.4\% with fluence values, which can be understood as the He$^{+}$ irradiation  intermixing the interfaces of [Co/Ni]/Cu and Cu/[Co/Pd].\cite{Devolder2000, Fassbender2004} The GMR maintains a value of 0.4\% at the highest fluence. 

\begin{figure}
\centering
\includegraphics[width=3.4in]{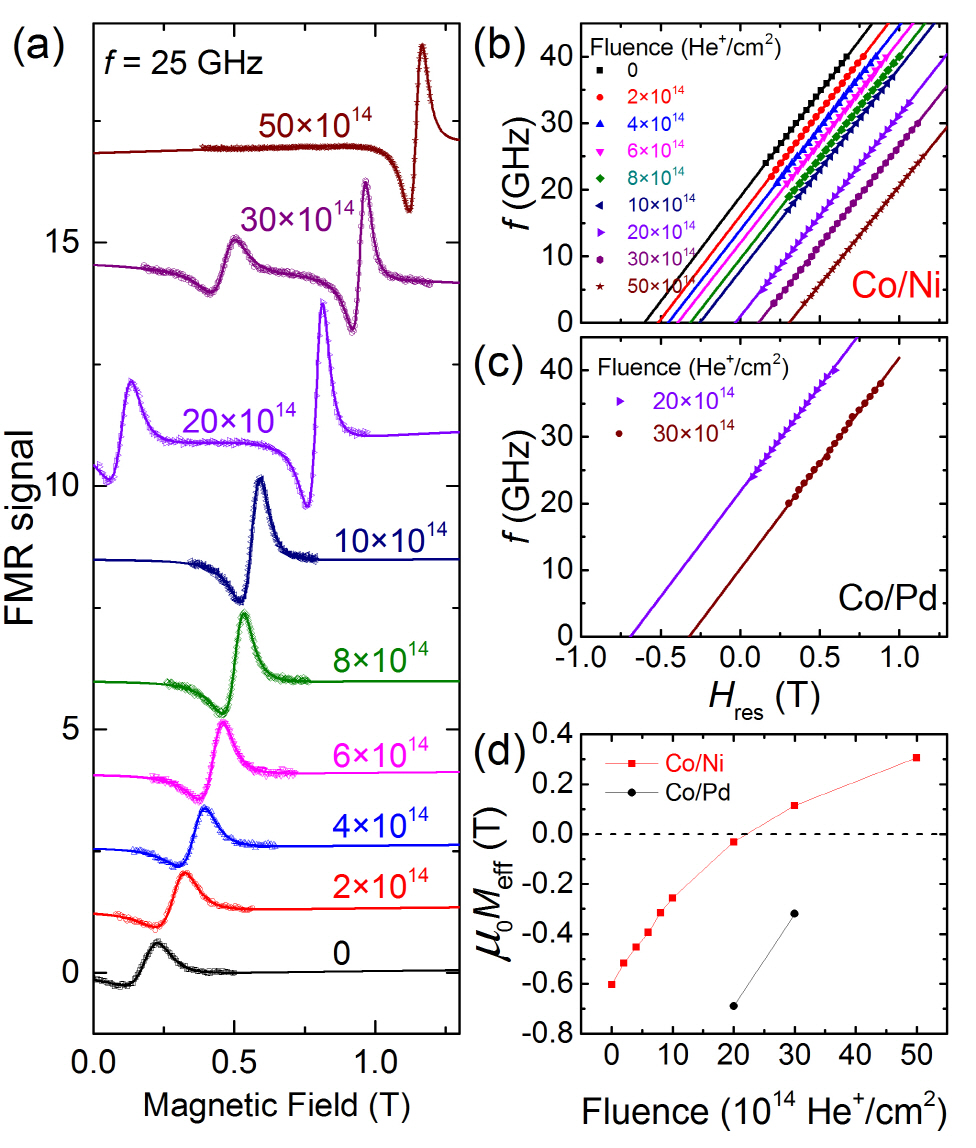}
\caption{\label{fig3} (a) Typical FMR spectra at 25 GHz. (b) and (c) Resonance field as a function of frequency for Co/Ni and Co/Pd, respectively. Solid lines are fits with the Kittel equation. (d) Effective magnetization extracted from FMR for Co/Ni and Co/Pd.
}
\end{figure}

We now turned our attention to the magnetodynamical properties of He$^{+}$-irradiated PSVs by conducting OOP FMR measurements. Figure~\ref{fig3}(a) shows the typical signal for different fluences at a frequency $f$~=~25~GHz. All spectra are fitted with a sum of symmetrical and antisymmetrical Lorentzian derivates,\cite{Woltersdorf2004, Yin2015} as shown by the solid lines in Fig.~\ref{fig3}(a). For fluences from 0 to 10$\times10^{14}$~He$^{+}$/cm$^{2}$, only one resonance peak appears. One more peak then appears  for $F$~=~20 and 30$\times10^{14}$~He$^{+}$/cm$^{2}$; again, one peak occurs for $F$~=~50$\times10^{14}$~He$^{+}$/cm$^{2}$. By fitting all  the data at different frequencies, we plotted the resonance field $H_{\mathit{res}}$ as a function of $f$ in Fig.~\ref{fig3}(b, c). We already know that the PMA of our Co/Ni is weaker than Co/Pd;\cite{Chung2013, Nguyen2011} we thus identify the single peak for the lower fluences with Co/Ni, and conclude that the Co/Pd resonance peaks are beyond the measured %field and 
frequency ranges because of its stronger PMA. However, as the fluences increase, the PMA of Co/Pd decreases, and the Co/Pd peaks can be observed for 20 and 30$\times10^{14}$~He$^{+}$/cm$^{2}$. For 50$\times10^{14}$~He$^{+}$/cm$^{2}$, the Co/Pd peak might be too weak to detect. We then fit the $H_{\mathit{res}}$ with the OOP Kittel equation,\cite{Kittel1949}
\begin{equation}
%\begin{split}
\label{eq1}
f=\frac{\gamma \mu _{0}}{2\pi }(H_{\mathit{res}}-M_{\mathit{eff}}),
%\end{split}
\end{equation}

where $\mu _{0}$ is the permeability of free space and $M_{\mathit{eff}}$ is the effective magnetization. $\gamma /2\pi$ is the gyromagnetic ratio, obtained by fitting with Eq.~\ref{eq1}; it is independent of fluences and equals 30.6 and 31.5~GHz/T for Co/Ni and Co/Pd, respectively. The extracted $\mu _{0}M_{\mathit{eff}}$ values are plotted in Fig.~\ref{fig3}(d). As the  fluence increases from zero to its highest value, $\mu _{0}M_{\mathit{eff}}$ of Co/Ni increases from -0.60 to 0.31 T. The negative value of $\mu _{0}M_{\mathit{eff}}$ at low fluences implies that the PMA is sufficient to overcome the demagnetizing energy, and hence the easy axis is normal to the film plane; so is that of Co/Pd (All-PMA). Interestingly, we observed that the $M_{\mathit{eff}}$ of Co/Ni changes sign to positive at 30$\times10^{14}$~He$^{+}$/cm$^{2}$, which indicates that Co/Ni is IP, while that of Co/Pd remains negative (OOP), %thus keeping OOP (orthogonal), 
in agreement with the results of HL in Fig.~\ref{fig1}(b). At  sufficiently high fluences (50$\times10^{14}$~He$^{+}$/cm$^{2}$), the PMA is reduced to the point where the demagnetization field dominates and the easy axes of both Co/Ni and Co/Pd lie in the film plane, %(all-IMA), 
as confirmed by the positive $\mu _{0}M_{\mathit{eff}}$ value of Co/Ni in Fig.~\ref{fig3}(c) and the only clear switching in Fig.~\ref{fig1}(c). The saturation magnetization $M_{s}$ is calculated from AGM data and exhibits no clear dependence on fluence for either Co/Ni or Co/Pd, with $\mu _{0}M_{s}$~=~1.0~$\pm $0.1~T and 1.2 $\pm $0.1~T, respectively. We thus claim that the anisotropy field $H_{k}$ shows a decreasing trend with fluence, since $M_{\mathit{eff}}$~=~$M_{s}$~-~$H_{k}$. We have hence demonstrated that irradiation with He$^{+}$ allows us to tune the magnetic structures from all-PMA, through orthogonal, to all-easy-plane.

\begin{figure}
\centering
\includegraphics[width=3.4in]{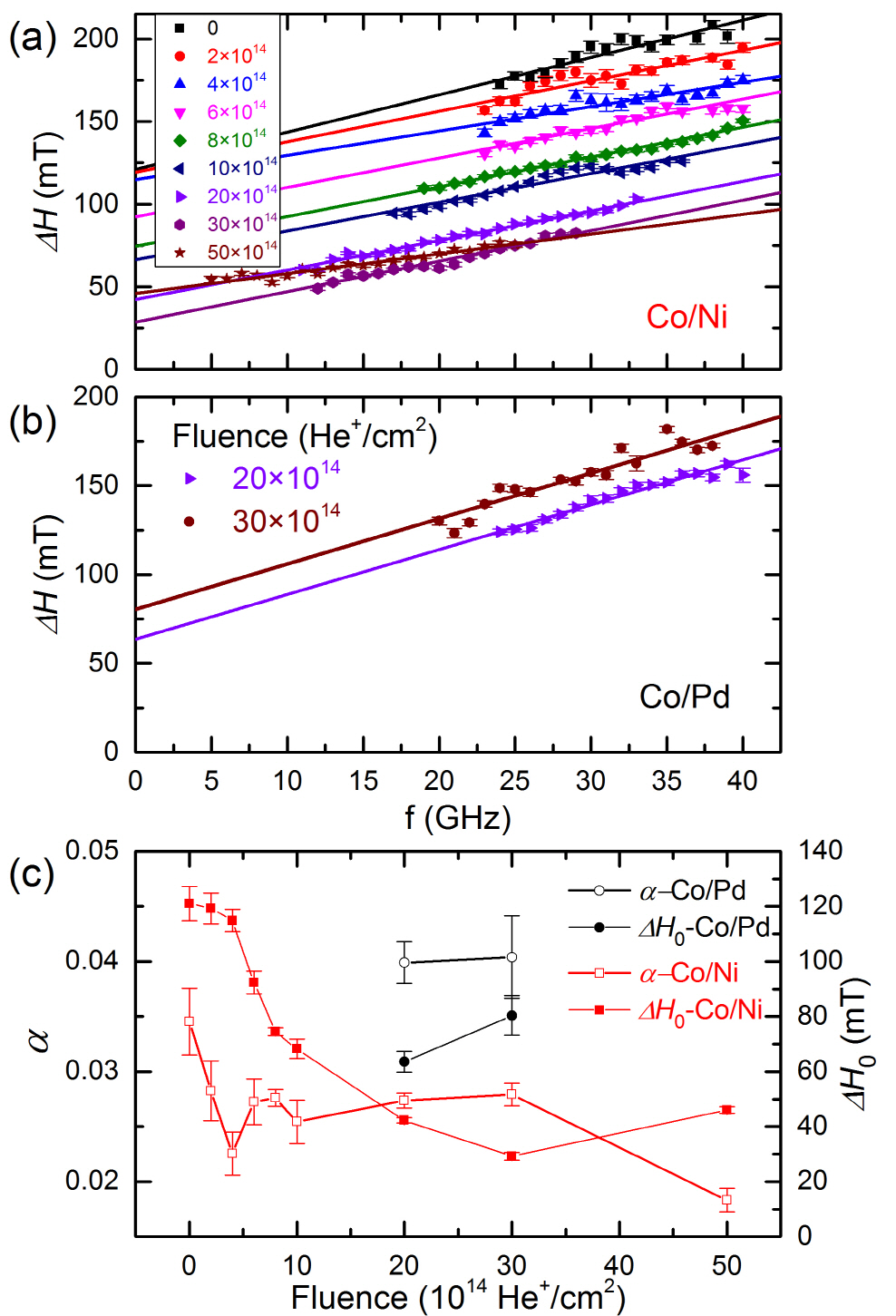}
\caption{\label{fig4} (a) and (b) Linewidth $\Delta H$ as a function of frequency for Co/Ni and Co/Pd, respectively. (c) Damping $\alpha$ and inhomogeneous linewidth $\Delta H_{0}$ as a function of fluence.
}
\end{figure}
In Figs.~\ref{fig4}(a, b), we extracted the full width at half maximum  (FWHM) linewidth $\Delta H$ by fitting the FMR spectra for Co/Ni and Co/Pd. The FMR linewidth contribution normally originates from the intrinsic Gilbert linewidth $\Delta H_{G}$,\cite{Gilbert2004, Brataas2008} inhomogeneous broadening $\Delta H_{0}$,\cite{Platow1998} and two-magnon scattering $\Delta H_{\mathit{TMS}}$.\cite{McMichael2003, Jiang2017} The Gilbert contribution, produced by the intrinsic spin-orbit coupling of the ferromagnetic materials, is proportional to the frequency $f$: $\Delta H_{G} \propto f$. Inhomogeneous broadening, independent of $f$, relies on sample inhomogeneity, which is probably associated with the local variation of $M_{\mathit{eff}}$, specifically the existence of interface roughness (see below). The two-magnon scattering, valid for defects as scattering centers in ferromagnets, is a process in which the $k$~=~0 magnon excited by FMR is scattered into degenerate magnon states with wave vectors $k$~$\neq $~0. Since there is no contribution of $\Delta H_{\mathit{TMS}}$ when the applied field is normal to the films,\cite{Jiang2017, Beaujour2009} we ignore the contribution of two-magnon scattering in our OOP FMR measurements. The linewidth can then be described as
\begin{equation}
%\begin{split}
\label{eq2}
\Delta H=\Delta H_{0}+\Delta H_{G}=\Delta H_{0}+\alpha \frac{4\pi }{\mu _{0}\gamma }f,
%\end{split}
\end{equation}
where $\alpha $ is damping constant. By fitting the linewidth with Eq.~\ref{eq2} in Figs.~\ref{fig4}(a, b), the extracted $\Delta H_{0}$ and $\alpha $ are shown in Fig.~\ref{fig4}(c). The damping of the Co/Ni layer is found to first improve substantially at low fluence (4$\times10^{14}$~He$^{+}$/cm$^{2}$), and then remain almost constant ($\alpha = 0.025 - 0.027$) at higher fluences, where it is still lower than the non-irradiated value. The improvement in the damping of Co/Ni (by a factor of two) may result from the intermixing of Co and Ni layers, ultimately becoming an alloy. The literature\cite{Schoen2017} has reported that the damping of Co ($\alpha _{Co}$~=~0.005) and Ni ($\alpha _{Ni}$~=~0.028) single layers is much lower than, and respectively comparable to, our Co/Ni multilayers. The damping of the alloy of Co$_{1-x}$Ni$_{x}$ has been investigated in Ref.~\onlinecite{Schoen2017}, where the damping shows a monotonic decrease with a decreasing in Ni concentration. For our He$^{+}$-irradiated Co/Ni multilayers, the intermixing of the Co and Ni layers by He$^{+}$ collision is analogous to the case of the alloying of Co and Ni. This could be one of the explanations for the reduction of damping with increasing fluence. The inhomogeneous broadening $\Delta H_{0}$ of the Co/Ni layer shows a dramatic drop after irradiation, and similar behavior has been reported for irradiated Co/Ni.\cite{Beaujour2009, Beaujour2011} The soft ion-induced intermixing may average out the role of defects at interfaces (interface roughness) by inducing a continuous Co/Ni alloy.\cite{Cayssol2005} As a result, the distribution of magnetic anisotropy is reduced leading to a decrease of the inhomogeneous broadening. on the other hand, The inhomogeneous linewidth is also associated with the grain sizes and proportional to the anisotropy field, as proposed in Ref.~[\onlinecite{McMichael2003}], which is consistent with our experimental observations. As we indeed observed, the anisotropy field $H_{k}$ and the inhomogeneous linewidth decrease as fluence increases. Regarding Co/Pd, the damping seems unaffected ($\alpha = 0.04$), and the inhomogeneous broadening is slightly larger for the two measurable fluences in Figs.~\ref{fig4}(b, c). To better understand the effect on Co/Pd, more detailed studies are needed, which is beyond the scope of this paper. The damping and inhomogeneous broadening of Co/Ni were improved simultaneously, which is critical for the free layer of STNOs, and suggests lower threshold currents and more uniform films. These parameters, however, are not as important for the fixed Co/Pd layer as for the free layer.
%\section{\label{sec:level1}Conclusions}

In conclusion, we investigated the controlled magnetic properties of [Co/Pd]/Cu/[Co/Ni] PSVs by He$^+$ irradiation. By performing hysteresis loop and magnetoresistance measurements, the coercivities of Co/Ni and Co/Pd showed a continuous reduction with increasing fluence, just like the GMR. FMR results showed that the perpendicular anisotropy field $H_{k}$ is progressively decreased by He$^{+}$ irradiation. This could result from the soft ion-induced interface mixing and strain relaxation. By precisely controlling the fluence, the remanent magnetic states can be adjusted from OOP to IP for both Co/Ni and Co/Pd, which allow us to achieve magnetic structures ranging from all-perpendicular ($\uparrow\uparrow$), through orthogonal ($\rightarrow\uparrow$), to all in-plane ($\rightrightarrows$). In addition, the damping and inhomogeneous broadening are improved simultaneously for Co/Ni, which benefits STNOs and STT-MRAM applications.

%(\textcolor{red}{Acknowledgments}) 
We gratefully acknowledge financial support from the China Scholarship Council (CSC), the Swedish Foundation for Strategic Research (SSF), the Swedish Research Council (VR), and the Knut and Alice Wallenberg Foundation (KAW). This work was also supported by the European Research Council (ERC) under the European Community's Seventh Framework Programme (FP/2007–2013)/ERC Grant 307144 ``MUSTANG.'', the French ANR project ELECSPIN number ANR-16-CE24-0018-02 and the Labex Nanosaclay.

\bibliography{main}

\end{document}